\documentclass[journal]{IEEEtran}
\usepackage{graphicx}          
\usepackage{amsmath}
\usepackage{tabularx}
\usepackage{amsfonts}
\usepackage{url}
\usepackage[table,xcdraw]{xcolor}
\usepackage{cite}
\usepackage{verbatim}
\usepackage{times,epsfig}
\usepackage{makecell}
\usepackage{psfrag}
\usepackage{subfigure}
\usepackage{stfloats}
\usepackage{setspace}
\usepackage{amssymb}
\usepackage{multirow}
\usepackage{colortbl}
\usepackage[justification=centering]{caption}
\usepackage{fancyhdr}
\usepackage{utfsym}
\usepackage{makecell}
\usepackage{booktabs}
\usepackage{tikz}
\usepackage{threeparttable}
\newcommand*\emptycirc[1][1ex]{\tikz\draw (0,0) circle (0.1);} 
\newcommand*\halfcirc[1][1ex]{%
	\begin{tikzpicture}
	\draw[fill] (0,0)-- (90:0.1) arc (90:270:0.1) -- cycle ;
	\draw (0,0) circle (0.1);
	\end{tikzpicture}}
\newcommand*\fullcirc[1][1ex]{\tikz\fill (0,0) circle (0.1);} 
\usepackage{hyperref}

\usepackage{enumitem}
\newlist{steps}{enumerate}{1}
\setlist[steps, 1]{label = Step \arabic*:}

\begin{document}
	\pagenumbering{arabic}
	
	\title{Adversarial Attacks and Defenses for Semantic Communication in Vehicular Metaverses}

	\author{Jiawen Kang, Jiayi He, Hongyang Du, Zehui Xiong*, Zhaohui Yang, Xumin Huang, Shengli Xie

	\thanks{Jiawen~Kang and Jiayi He are with the School of Automation, Guangdong University of Technology, Guangzhou 510006, China, and also with the Guangdong-Hong Kong-Macao Joint Laboratory for Smart Discrete Manufacturing, Guangzhou 510006, China. Xumin Huang is with the School of Automation, Guangdong University of Technology, Guangzhou 510006, China, and also with the 111 Center for Intelligent Batch Manufacturing Based on IoT Technology, Guangzhou 510006, China. Shengli Xie is with the School of Automation, Guangdong University of Technology, Guangzhou 510006, China, and also with the Guangdong Key Laboratory of IoT Information Technology, Guangzhou 510006, China. Hongyang Du is with the School of Computer Science and Engineering, Nanyang Technological University, Singapore. Z. Yang is with the College of Information Science and Electronic Engineering, Zhejiang University, Hangzhou 310007, China, and Zhejiang Provincial Key Lab of Information Processing, Communication and Networking (IPCAN), Hangzhou 310007, China, and also with Zhejiang Laboratory, Hangzhou 31121, China. Zehui~Xiong is with the Pillar of Information Systems Technology and Design, Singapore University of Technology and Design, Singapore 487372, Singapore. (\textit{*Corresponding author: Zehui Xiong})}
	}
	\maketitle
	\pagestyle{headings}

	\begin{abstract}
    
        For vehicular metaverses, one of the ultimate user-centric goals is to optimize the immersive experience and Quality of Service (QoS) for users on board. Semantic Communication (SemCom) has been introduced as a revolutionary paradigm that significantly eases communication resource pressure for vehicular metaverse applications to achieve this goal. SemCom enables high-quality and ultra-efficient vehicular communication, even with explosively increasing data traffic among vehicles. In this article, we propose a hierarchical SemCom-enabled vehicular metaverses framework consisting of the global metaverse, local metaverses, SemCom module, and resource pool. The global and local metaverse are brand-new concepts from the metaverse's distribution standpoint. Considering the QoS of users, this article explores the potential security vulnerabilities of the proposed framework. To that purpose, this study highlights a specific security risk to the framework's SemCom module and offers a viable defense solution, so encouraging community researchers to focus more on vehicular metaverse security. Finally, we provide an overview of the open issues of secure SemCom in the vehicular metaverses, notably pointing out potential future research directions.
    
	\end{abstract}

	
	\section{Introduction}
	
    With the advances of the Internet of Things and Artificial Intelligence, metaverse technology, regarded as the next-generation Internet, is rapidly emerging to build a virtual-physical integrated world with fully immersive and personalized experiences for users in many scenarios and applications. In particular, vehicular networks are revolutionizing towards vehicular metaverses that have vast potential to provide diverse, immersive, and personalized in-vehicle entertainment/services for both drivers and passengers \cite{9880566}.
    In vehicular metaverses, vehicle data are mainly divided into two categories: i) static data including background information on roadside infrastructures and road networks, and ii) dynamic data involving information about pedestrians, moving vehicles, traffic flow, and so on.
    To support time-sensitive vehicular metaverse services (e.g., AR navigation), a large number of dynamic data is required for real-time updates of the vehicular physical-virtual world to avoid service quality degradation, while conventional communication based on Shannon Information Theory cannot support efficient communications and massive connectivity for vehicular metaverses due to limited wireless communication resources.
		
	Recently, Semantic Communication (SemCom) has been introduced as a revolutionary paradigm that breaks through the bandwidth bottleneck of classical communications, significantly improving communication efficiency by leveraging deep learning models to only transmit critical information sufficient for the receivers, thus dramatically reducing the number of bits transmitted.
	A deep learning-enabled end-to-end SemCom system with Deep Neural Networks (DNNs) as semantic codecs has been proposed in~\cite{xie2021deep}. It is a promising solution to integrate SemCom into vehicular metaverses for enabling real-time transmission of vast data used to sustain the metaverse services. In vehicular metaverses, vehicles only send data requests and then download lots of static and dynamic data from edge servers in the way of SemCom with unprecedented communication efficiency in comparison to traditional communication manners.
	 
	Although semantic communication plays an indispensable role in vehicular metaverses, SemCom-enabled vehicular metaverses are still in their infancy. There are many challenges to be resolved for realizing its potential, especially privacy and security issues. On the one hand, an eavesdropper can infer vehicle location and driving habits by analyzing communication data. Even though privately trained codecs provide SemCom with a natural barrier to being eavesdropped, this barrier does not work well in vehicular metaverses. This is because there exist similar and stationary communication tasks in neighbor vehicles and then the eavesdropper may obtain similar or even identical decoders to recover data from the wireless channels. The researches on \cite{luo2022encrypted,tung2022deep,HongyangDu2022RethinkingWC} are committed to private communications. On the other hand, DNNs are vulnerable to attacks (e.g., poisoning attacks and adversarial attacks). Such weaknesses naturally exist in deep learning-based SemCom systems. Attackers can easily apply typical deep learning attacks to the SemCom systems and directly reduce the task/model accuracy ~\cite{hu2022robust}. Non-malicious perturbations may cause performance degradation of semantic communication systems as well~\cite{XiangPeng2022ARD}. 

    In particular, the existing work ignores adversarial attacks of SemCom in vehicular metaverses. For adversarial attacks, the authors in~\cite{szegedy2013intriguing} find that non-random perturbation on the test sample can arbitrarily manipulate the output of neural networks. In general, the attacker first obtains the structure and parameters of the target model. The adversarial samples are then initialized with the original data. After several training iterations, the elaborate adversarial samples similar to the original data are generated by maximizing the loss of the semantic encoder. Finally, the accuracy degradation of the target model is achieved in the face of adversarial samples. Hu et al. in ~\cite{hu2022robust} demonstrate that adversarial attacks are still feasible in the deep learning-enabled SemCom system. For vehicular metaverses, the adversarial attacks increase the probability of receiving wrong information, even causing traffic accidents if receivers are misled to take dangerous actions according to wrong information~\cite{hu2022robust,luo2022frequency}.
    
    To fully understand the risks of adversarial attacks for SemCom-enabled vehicular metaverses, in this article, we study the adversarial attacks and the corresponding defenses against these attacks. More specifically, we first design a hierarchical vehicular metaverse framework with a global metaverse and multiple local metaverses~\cite{hashash2022towards}. A new adversarial attack model and its defense scheme for this framework are then proposed, respectively. We discuss several security challenges of SemCom-enabled vehicular metaverses. The main contributions are summarized as follows:
	
	\begin{itemize}
	    \item According to vehicle data characteristics and service requirements, we design a new and hierarchical SemCom-enabled vehicular metaverse framework mainly consisting of a global metaverse, multiple local metaverses, and resource pools. This framework not only significantly reduces data transmission delay, but also efficiently utilizes the computing and storage resources of edge servers.
		\item We propose a novel adversarial attack called Semantic Noise Attack (SNA) to generate adversarial samples by adding semantic noise and design a new defense scheme using the Semantic Distance Minimization (SDM) mechanism to weaken the adversarial samples, but almost without sacrificing transmission accuracy. 
		\item For use cases, we present two SemCom systems with different semantic encoders on traffic sign classification and license plate recognition, respectively. We examine the effects of SNA on the SemCom systems and evaluate the robustness of the defense scheme based on SDM in vehicular metaverses. The numerical results indicate that the proposed defense scheme significantly reduces the success rate of the attacks.
	\end{itemize}
 \captionsetup[figure*]{labelfont=bf,textfont=normalfont,singlelinecheck=off,justification=raggedright}
\begin{figure*}[h]
\centering
\includegraphics[width=0.95\textwidth]{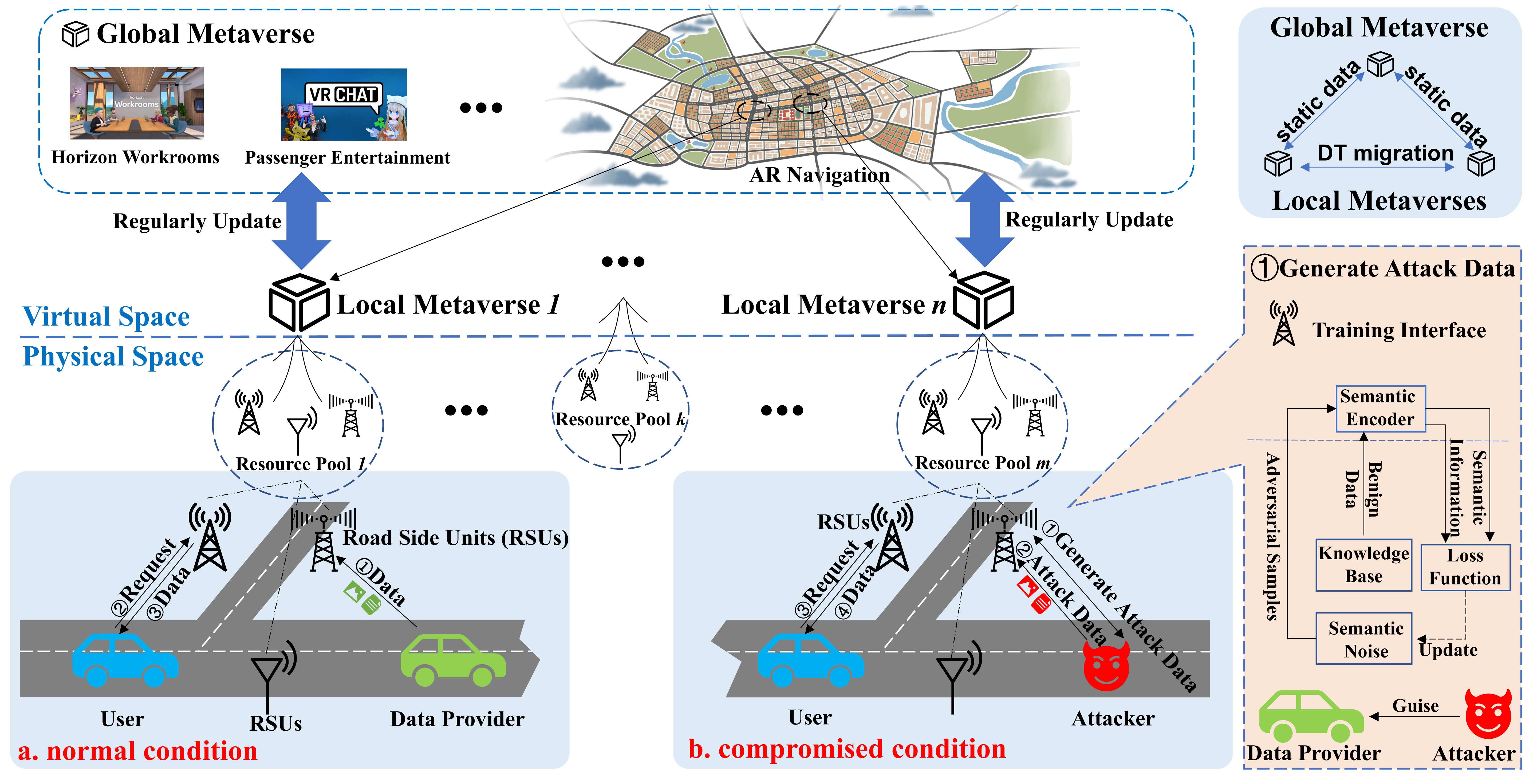}
\caption{The hierarchical vehicular metaverse architecture. Parts a and b show the normal service and attack flow respectively; The orange subfigure on the right-hand side depicts the steps involved in generating attack examples, while the blue subfigure illustrates the communication between the global metaverse and the local metaverses.}

\label{fig1}
\end{figure*}	
	
\section{Semantic Communication for Vehicular Metaverses}	
    For vehicular metaverses, sensing data is divided into static and dynamic data based on the properties in the real world. Static data is the background information from the physical space, such as road sensors, roadside infrastructures, and the topology network of roads.
    Dynamic data is the ``add-in" information from dynamic entities in the physical space including vehicles, pedestrians, traffic signs, etc. Actually, the dynamic data is used to update the digital twins of dynamic entities in the virtual space. Compared with other metaverse services, vehicular metaverses have stricter latency requirements to ensure user immersion and improve the QoS.
	
	\subsection{Semantic Communication and Vehicular Metaverses}

    As shown in Fig.~\ref{fig1}, a hierarchical vehicular metaverse consists of a global metaverse, \textit{n} local metaverses, and \textit{m} resource pools. More specifically,

    \begin{itemize}
    	\item \textit{Global Metaverse}: a giant metaverse deployed over the cloud by Metaverse Service Providers (MSPs) for management globally. The global metaverse is a replica of the physical world in the virtual space, which contains both static data and dynamic data within an entire city or even wider area, e.g., a global city map with abundant traffic-related information in the navigation applications. Instead of massively communicating directly with users, it was updated periodically by the data from all the local metaverses. 
    	\item \textit{Local Metaverses}: as a component of the global metaverses, a local metaverse could be  the metaverse mapping to a small-scale area, like  a street block or a road intersection. The local metaverses are established  on edge servers and are closely connected to nearby users, and thus the  vehicles can communicate directly with their local metaverses to obtain real-time metaverse  services without large latency ~\cite{hashash2022towards}.
    	\item \textit{Resource Pools}: the virtual resource pool consisting of physical entities with sufficient resources in the local metaverse, e.g., roadside units acting as edge servers. The edge servers in different regions establish  a resource pool for their common local metaverse, which includes computing, storage, and bandwidth resources to support the metaverse services for users.
    \end{itemize}

    The hierarchical framework has several new designs and thus obtains multiple advantages. First, proximity to the users allows the framework to avoid long-distance communication, and thus reduce communication time. Users only need communication with the closest node in the resource pool to access the services. Second, the framework trims the budget for deploying dense sensor nodes in physical space. The strategy for updating and collecting data does not require massive sensors~\cite{hashash2022towards}. Third, the framework efficiently utilizes the storage and network resources of servers and vehicles. Both the global and local metaverses only hold indispensable data to maintain normal service. The vehicles only download the data of the local metaverses corresponding to their location. Finally, the SemCom in the framework reduces the number of bits that need to be transmitted. Hence, the effects of channel instability caused by vehicle movement are almost mitigated.

    In vehicular metaverses, for AR navigation, the MSP collects massive static data in a low-cost way (e.g., from satellites) and deploys them on the global metaverse. As shown in \ref{fig1}, the static data at each intersection is regularly updated to corresponding resource pools to build local metaverses. The resource pools collect dynamic data from data providers in the physical world to keep updating digital twins. Meanwhile, when a physical entity (e.g., a vehicle) moves from one local metaverse to another one, its digital twin may migrate to the new local metaverse according to its location. Note that all the changes in the static collected data will be updated to the global metaverse. Rather than encapsulating an entire city's dynamic and static data in one metaverse, the local metaverses split the data, thus relieving the communication and computing pressure of the cloud and resource pools and shortening the distance between users and data. Vehicles can interact with edge servers in the resource pools to access metaverse services.

    Before entering the coverage of a local metaverse, each user, i.e., vehicle, initiates a service request to the edge server. Once a user request is received, the edge server sends static information from the local metaverse to the vehicle. Meanwhile, the required dynamic data are constantly invoked and sent to the vehicle, allowing the driver to glimpse navigation and driving prompts. Congested wireless resources cannot support the transmission of such large amounts of data with low latency. To reduce latency, the edge servers only transmit key features of data to the vehicles by SemCom.

    In vehicular metaverses, vehicles do not need to recover the source data, but only need to complete pragmatic tasks (e.g., traffic sign recognition and pedestrian detection). Given that the tasks in vehicular metaverses are relatively fixed, we design task-oriented SemCom systems~\cite{xie2022task} to improve efficiency. The task-oriented SemCom aims to extract more condensed semantic information related to the task of the receiving end. The receiver no longer needs a complex semantic decoder to recover the source data but completes the task directly based on the semantic information with a simple model. Coincidentally, this is consistent with the computational power distribution between resource pools and vehicles.

    We consider that each edge server has a dedicated semantic encoder for every task. Due to the heterogeneity of hardware performance, driving habits, and knowledge base, each vehicle has a customized task model. Hence, users need to train their models in conjunction with edge servers before enjoying the services. The edge servers open interfaces at night or during idle hours for user training. The interface mechanically extracts the semantic information of the original data given by the user and sends it back. Therefore, the user can train their task model by semantic information and the labels of original data.
    
    After completing the training, the user can access the metaverse to enjoy application services. Three kinds of participants are involved in a service process as follows: 

    \begin{itemize}
    	\item \textit{User}: In vehicular metaverses, the vehicle usually acts as a user. The proposed framework and all the mechanics are designed to make the user obtain a safe, reliable, comfortable, and immersive experience.
    	\item \textit{Edge Server}: The intermediary between the user and the data provider. The edge servers collect data from the data provider, store it temporarily and send it to the user when receiving a request.
    	\item \textit{Data Provider}: Any sensing node that wants to make money by selling data (e.g., some users, roadside cameras). A data provider located near an intersection contingently provides data to the edge servers in the corresponding resource pool.
    	
    \end{itemize}

    Signal lights, traffic signs, pedestrians, and other information are captured by the data providers’ sensors and sent to the edge server. Based on SemCom, the edge server sends the data to the user with a graceful delay after the user initiates the request. By the task models, the driver can timely see the arrow guides and virtual traffic sign entities in front of the windshield, which can enhance the user's perception of road conditions. This driving paradigm can greatly improve the safety and comfort of driving, especially when the view is blocked by trucks or buildings. Passengers can also use this system to play metaverse games or work online.

    \begin{table*}[t]
        \centering
        \caption{ Existing Works Comparison.}
        \label{tab:works}
        \begin{threeparttable}
        \begin{tabular}{lcccccc}
            \hline
            \textbf{Method} & \textbf{Background} & \textbf{Attack} & \textbf{Defense} & \textbf{Dataset} & \textbf{Modal} & \textbf{Model} \\ \hline
            ESCS~\cite{luo2022encrypted} & \emptycirc & Eavesdrop & Encryption & Europarl & Text & Transformer\\ 
            VQ-VAE~\cite{hu2022robust} & \emptycirc & Adversarial attack & Mask + Codebook & \thead{CIFAR-10 \\Cars196\\ImageNet} & Image & ViT\\
            R-DeepSC~\cite{XiangPeng2022ARD} & - & Semantic noise & \thead{Calibrating \\ Augment training set} & Europarl & Text & Transformer\\
            DeepJSCEC~\cite{tung2022deep} & \emptycirc & Eavesdrop & Encryption & CIFAR-10 & Image & DNNs\\
            \color{red} SNA\&SDM & \fullcirc & \color{red}SNA & \color{red}SDM & \color{red}\thead{GTSRB\\CCPD} & \color{red}Image & \color{red}\thead{ResNet\\LPRNet}\\
            \hline
        \end{tabular}
        \begin{tablenotes} 
            \item The \emptycirc~represent white box attack, while the \fullcirc~means black box setting. In the white-box setting, the attacker possesses knowledge of the structure or parameters of the semantic codec model. In contrast, the attacker in the black-box setting is ignorant and typically attacks by probing the model's input and output.
        \end{tablenotes} 
        \end{threeparttable} 
        \end{table*}

	\subsection{Security Challenges of Semantic Communication}
	
	SemCom, as a promising communication paradigm for the 6G future era, shows good performance in various tasks. However, it is vulnerable to some well-designed attacks. Therefore, it is necessary to consider the possible threats and defend against these attacks. We summarize typical security challenges as follows:
	\begin{itemize}
		\item \textit{Eavesdropping Attack}: Eavesdropping is the behavior of decoding information in a physical channel in order to steal private information. In SemCom, the user-unique semantic codec acts as a barrier to eavesdroppers. Even though an eavesdropper obtains the transmitted semantic information, it cannot decode it. However, this barrier is not foolproof. When the SemCom system is widely used, users from the same communication link may have the same semantic decoder structure. Lu et al.~\cite{lu2020semantic} reveal the privacy issues of the SemCom system in the industrial Internet of Things. At the application layer, Luo et al.~\cite{luo2022encrypted} propose an adversarial training method with key pairs to prevent such eavesdropping. Tung et al.~\cite{tung2022deep} propose a joint source-channel and encryption coding scheme for wireless image transmission. At the physical level, Chorti et al.~\cite{chorti2022context} consider physical layer security as the essential issue in next-generation communication. Du et al.~\cite{HongyangDu2022RethinkingWC} consider this problem from a physical perspective and protect wireless communications against eavesdropping by exploring and utilizing the inherent features of the physical medium.
		\item \textit{Adversarial Attack}: By adding imperceptible semantic noise to the transmitted data, adversarial attacks make the SemCom system produce errors in the encoding stage or decoding stage. The authors in~\cite{hu2022robust} consider two types of attacks, one is on the sending end and the other one is on the receiving end. The former simulated the case where the attacker is the sender, while the latter simulated the case where the attacker initiates the attack through the channel. With all or part of the detailed information of the SemCom system, the attacker generates adversarial samples against the model. Xiang et al. ~\cite{XiangPeng2022ARD} use the calibrated network and expand the training set to counteract the semantic noise in the text transmission tasks.
		\item \textit{Poisoning Attack}: Poisoning attacks aim at manipulating the training process to degrade model performance or to insert backdoors. SemCom systems are trained based on shared semantic knowledge bases. The communication parties are likely to expand the training set from the third party to obtain a better SemCom codec. Some low-quality data and malicious data are inevitable, which may carry the wrong semantic information. Using data that contains false semantic information for training may lead to a degradation of performance in the SemCom system. This type of attack is called \textit{semantic data poisoning attack}~\cite{yang2022semantic}. Researches on poisoning attacks and their defenses in the SemCom system are still in the vacant stage. 
	\end{itemize}
 
	We summarize the existing work in Table I. Unlike prior works, we propose a black-box model, that is more realistic, and use datasets, that are consistent with vehicular metaverse scenarios. Moreover, we tailor semantic encoder models for different tasks.
	\section{Adversarial Attacks for Semantic Communication in Vehicular Metaverse}
	
	\subsection{Adversarial Attacks}

	Deep learning models are often targeted by adversarial attacks, where attackers generate adversarial samples that appear natural to human eyes but cause the model to produce incorrect outputs. Adversarial samples are typically created by adding specific noise that appears to be normal noise to clean samples. In addition to channel noise and attenuation, SemCom furthermore faces a special noise called \textit{semantic noise}~\cite{hu2022robust}. While channel noise is simulated to improve performance in real environments, semantic noise, which can subtly alter the meaning of data, is not typically considered. Semantic noise usually results from poor-quality data due to aging sensors and channel interference. By artificially adding semantic noise, attackers can efficiently produce adversarial samples, which cause a devastating blow to the SemCom system. We propose an attack based on semantic noise for vehicular metaverses.
	
	According to Section \uppercase\expandafter{\romannumeral2}, it is stated that edge servers are nodes that are semi-trusted and do not actively attempt to attack the system, but can still be deceived. These servers forward both accurate and inaccurate data to users. Typically, edge servers assess the reliability of data based on the reputation of their provider. However, this method fails when a provider is compromised or intentionally behaves in a trustworthy manner to gain credibility.
	
	The diagram in Fig.~\ref{fig1} illustrates how an attacker can impersonate a data provider and deceive well-behaved edge servers. Initially, the attacker collects natural data as if they were a legitimate provider. Based on the captured data, the attacker generates several adversarial samples through the SNA. At an opportune moment, the attacker submits these generated adversarial samples to the edge server. When a user requests information from the server, it extracts incorrect semantic details and sends them back to the user. This misinformation may lead users to make risky decisions. The process of generating adversarial samples through SNA is further explained in Fig.~\ref{fig1} and described below:
	
	\begin{itemize}
      \item \textit{Step 1}: The attacker gains access to the training interface offered by the edge server and initiates a training procedure.
      \item \textit{Step 2}: The attacker creates a group of adversarial samples using the benign data and then transmits both sets to the interface.
      \item \textit{Step 3}: The interface extracts semantic information using the semantic encoder and transmits it back to the attacker.
      \item \textit{Step 4}: In the received semantic information, the attacker calculates the maximum semantic distance to the benign data for every adversarial sample as the loss.
      \item \textit{Step 5}: Update adversarial samples based on the loss and jump to \textit{Step 2} until reaching the maximum iterations.
    \end{itemize}
	
	Using SNA, an attacker can calculate its loss by utilizing the interface provided by the edge server as environmental feedback, without knowing the structure and parameters of the semantic encoder. Edge servers aim to attract numerous data providers due to their requirement for dynamic data. As a result, detecting an infiltrating attacker is challenging, particularly when it attempts to enhance its reputation. It should be noted that the loss not only includes the steps necessary for executing an effective attack but also controls low-frequency perturbations~\cite{luo2022frequency} to minimize detection by human observers.
	
	Once the SNA is implemented, it brings incalculable losses to the MSP and users. For MSPs, the presence of an attacker means that pre-trained semantic encoders can no longer be used, resulting in financial loss. For users, incorrect semantic information can cause drivers and passengers to watch lower-quality virtual entities and thus reduce immersion. In some applications, incorrect semantic information can even further manipulate user behaviors. For example, in AR navigation, a stop sign contains the semantic information of going straight. The user continues to drive straight based on semantic information, thus causing an accident.

    \subsection{Defense for Adversarial Attacks}
    \captionsetup[figure]{labelfont=bf,textfont=normalfont,singlelinecheck=off,justification=raggedright}
    \begin{figure}
    \centering
    \includegraphics[width=0.45\textwidth]{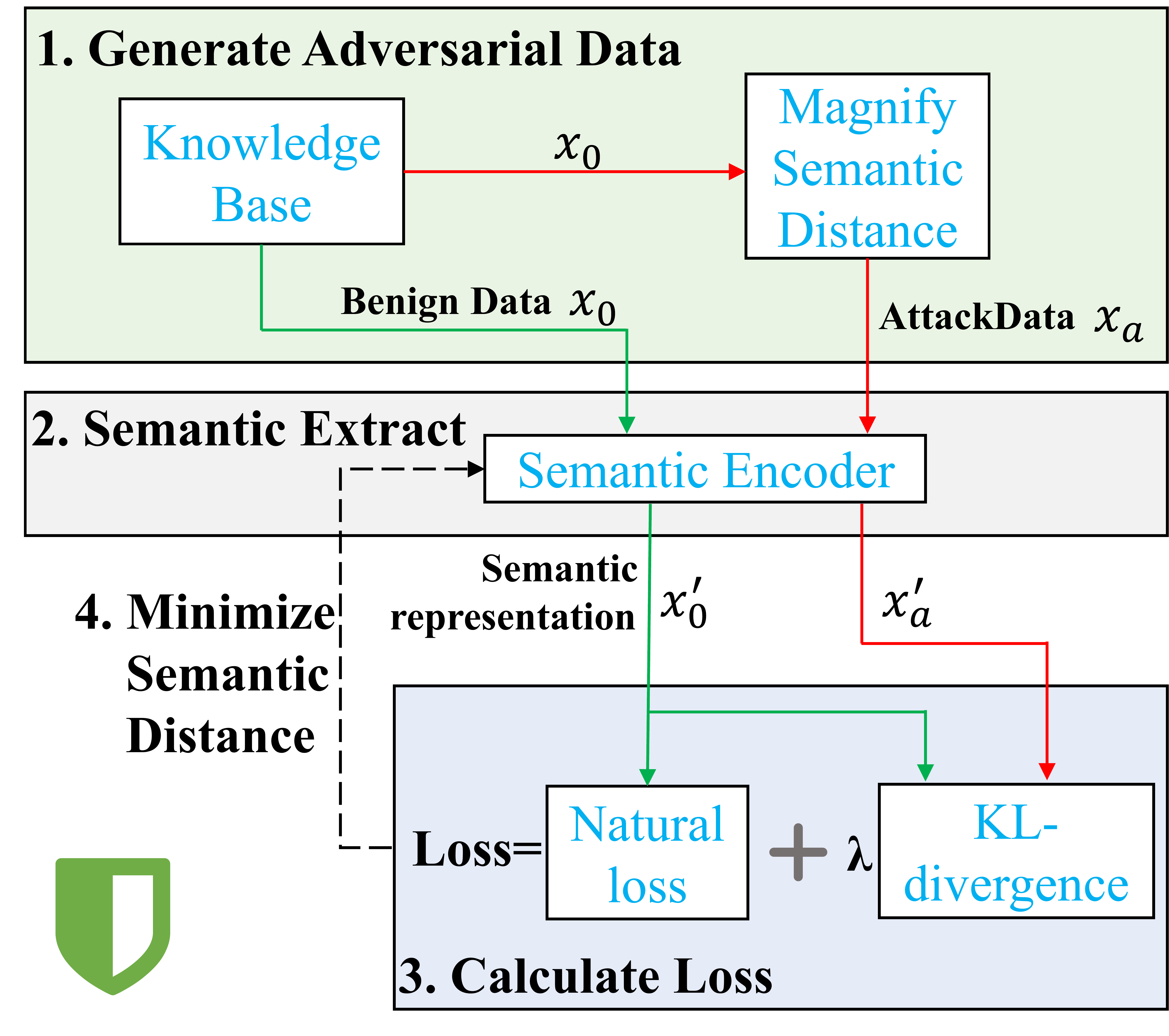}
    \caption{Semantic Distance Minimization Defense Process, where lines with the same color represent the same data stream, the dashed line indicates that the semantic encoder is updated according to the loss function.}
    \label{fig2}
    \end{figure}
    
    \begin{figure*}[h]
    \centering
    \includegraphics[width=0.95\textwidth]{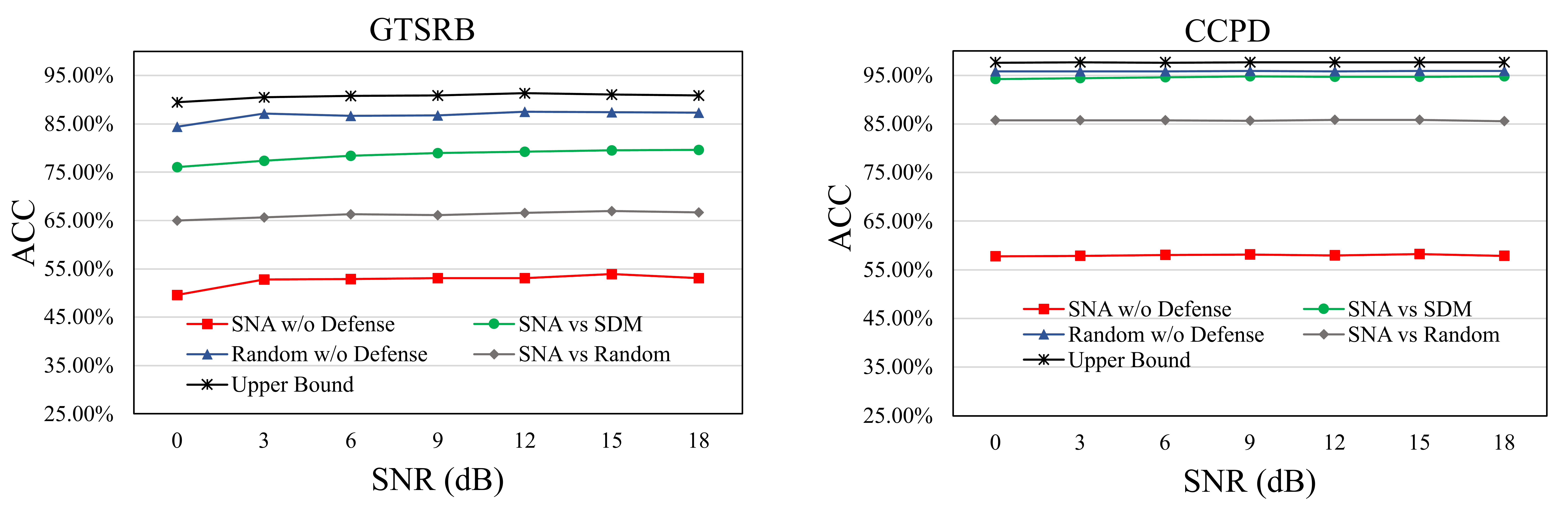}
    \caption{Accuracy versus the SNR on the adversarial test set with $l2\approx3.6$, where $a$ vs $b$ curve represents the $b$ target model attacked in $a$ attack mode. The random attack is to add stochastic noise to the source images.}
    \label{fig3}
    \end{figure*}
    As aforementioned, the SNA aims to modify semantic information by introducing semantic noise. In nature, semantic noise is the addition of disturbance in the efficient direction, which can change the position of data in the semantic space with the minimum disturbance. Generally speaking, one straightforward solution is to make the semantic encoder robust to semantic noise. Adversarial training is an attractive alternative but comes at the cost of natural accuracy. This article introduces the trade-off between natural and robust accuracy~\cite{zhang2019theoretically} into SemCom for vehicular metaverses to achieve SDM.

    Adversarial training involves generating online adversarial samples that result in maximum semantic loss and allows the model to learn their distribution. In SDM, as illustrated in Fig.~\ref{fig2}, online adversarial samples are generated for each data point through several iterations to maximize the distance from benign data. The KL divergence between the outputs of the semantic encoder is used to measure semantic distance. Both adversarial data and benign data are fed into the semantic encoder to calculate their semantic information.~\cite{zhang2019theoretically} The complete loss function includes two parts, natural loss, and robust loss. The robustness loss is determined by calculating the distance of extracted semantic information with a coefficient $\lambda$. Natural loss refers to the original mission's (e.g., classification) objective of improving model accuracy, while robust loss aims at ensuring that both adversarial samples and benign data have similar semantic information.

	\section{Case Study: Semantic Communication in Vehicular Metaverse}
	
	As a case study, we simulate a scenario to evaluate proposed attack methods and defense performance. In this scenario, an attacker deduces the user's location, produces adversarial samples, and provides them to RoadSide Units (RSUs) before the user accesses a local metaverse. To achieve this goal, we build SemCom systems in the above framework that simulate two tasks of AR navigation service in vehicular metaverses. The performance of the system performing traffic sign recognition and license plate recognition is measured. We conducted an SNA attack against these missions to assess the effectiveness of the attacks. Defensive SemCom systems are trained according to SDM, and their performance is re-evaluated. The same attack works on the new SemCom systems to test them. 

	\subsection{Simulation settings and Security Attacks}
	
    We consider an attacker capable to deduce the user location and provide adversarial samples to RSUs before the user enters the local metaverse. In this context, we develop SemCom systems to simulate two tasks related to AR navigation services in vehicular metaverses. Specifically, we design semantic codecs for each task, while the channel encoder and decoder are both with dense layers with different units. During training, Additive White Gaussian Noise (AWGN) is added to the channel with a Signal Noise Ratio (SNR) between 5 and 10~\cite{xie2021deep}, allowing the SemCom system to adapt for various real communication environments.
	
	\textit{1)} SemCom for traffic sign recognition: The GTSRB~$\footnote{https://benchmark.ini.rub.de/}$dataset consists of more than 50,000 images between $15*15$ to 250*250 pixels, divided into 43 classes. The pragmatic task of traffic sign recognition is completed in the communication process without the need to train the pragmatic function separately. In this simulation, the semantic encoder is a ResNet, while the receiver uses a fully connected layer to get the classification.
 
	\textit{2)} SemCom for license plate recognition: The CCPD~$\footnote{https://github.com/detectRecog/CCPD}$ dataset consists of more than 250,000 vehicle images from China for license plate detection and recognition. In this simulation, the pragmatic task is license plate recognition, so we pre-process the dataset according to the annotation, cutting the license plate part of its image. The semantic encoder is an LPRNet~$\footnote{https://arxiv.org/abs/1806.10447}$, while the receiver uses a greedy decoder to get the license plate information.
	
	As for the attack, to simulate the SNA in vehicular metaverses, we generate adversarial samples of all the test images and perform the corresponding pragmatic tasks through the SemCom system. The same attack is performed on the defensive model. We conducted an SNA attack against these missions to assess the effectiveness of the attacks. We set the same $L2$ norm for all attacks to ensure that performance is compared at the same attack strength~\cite{hu2022robust,luo2022frequency}.
	
	\subsection{Metrics and Defense in Semantic Communication }
	
	The robust SemCom models are trained with the aforementioned defense method for the two pragmatic tasks. During the generating online adversarial images phase, $10$ iterations are performed to maximize the semantic loss. In the training phase, we set $\lambda$ as $1$ in the traffic sign recognition task, which is $0.1$ in the license plate recognition task to achieve better defensive performance. The same attack works on the new SemCom systems to test them.
	
	The natural accuracy of the SemCom system on the test set is used to characterize its performance. For license plate recognition, only when all the symbols on the license plate are correctly recognized can be counted. We replace all images in the test set with the adversarial samples from the SNA and recalculate the accuracy, called \textit{robust accuracy}. To evaluate the effect of an attack, we use natural accuracy as an upper bound. The degree of decline in robust accuracy compared to the upper bound indicates the effectiveness of an attack. More reduction in accuracy means more successful attacks. We also calculate the accuracy for SDM SemCom systems to demonstrate defensive effects. On the contrary, lower accuracy cuts mean more successful defenses. In addition, the difference in natural accuracy between the original SemCom system and the SDM system demonstrates the defense overhead. These assessments are repeated in different SNR environments. 
	
	\subsection{Numerical Results}
		
		Figure \ref{fig3} displays the SNR to accuracy curves of different models when tested on adversarial samples. In the task of recognizing traffic signs, the SemCom model's accuracy is significantly reduced by $40\%$ due to the SNA method. This reduction in precision cannot ensure the normal operation of services and may even mislead drivers into making dangerous decisions based on incorrect road signs. As random channel noise is considered during model training, adding random disturbance through attack methods only causes a slight decrease in precision. The SDM SemCom system exhibits higher accuracy than the original system against SNA attacks. Additionally, random defense adds random disturbance to benign images instead of maximizing semantic loss to simulate an attack sample, resulting in weaker defensive effects compared to our proposed approach shown in Figure \ref{fig3}. This phenomenon still exists with global deviation only in license plate recognition tasks.
		
\begin{figure}
\centering
\includegraphics[width=0.5\textwidth]{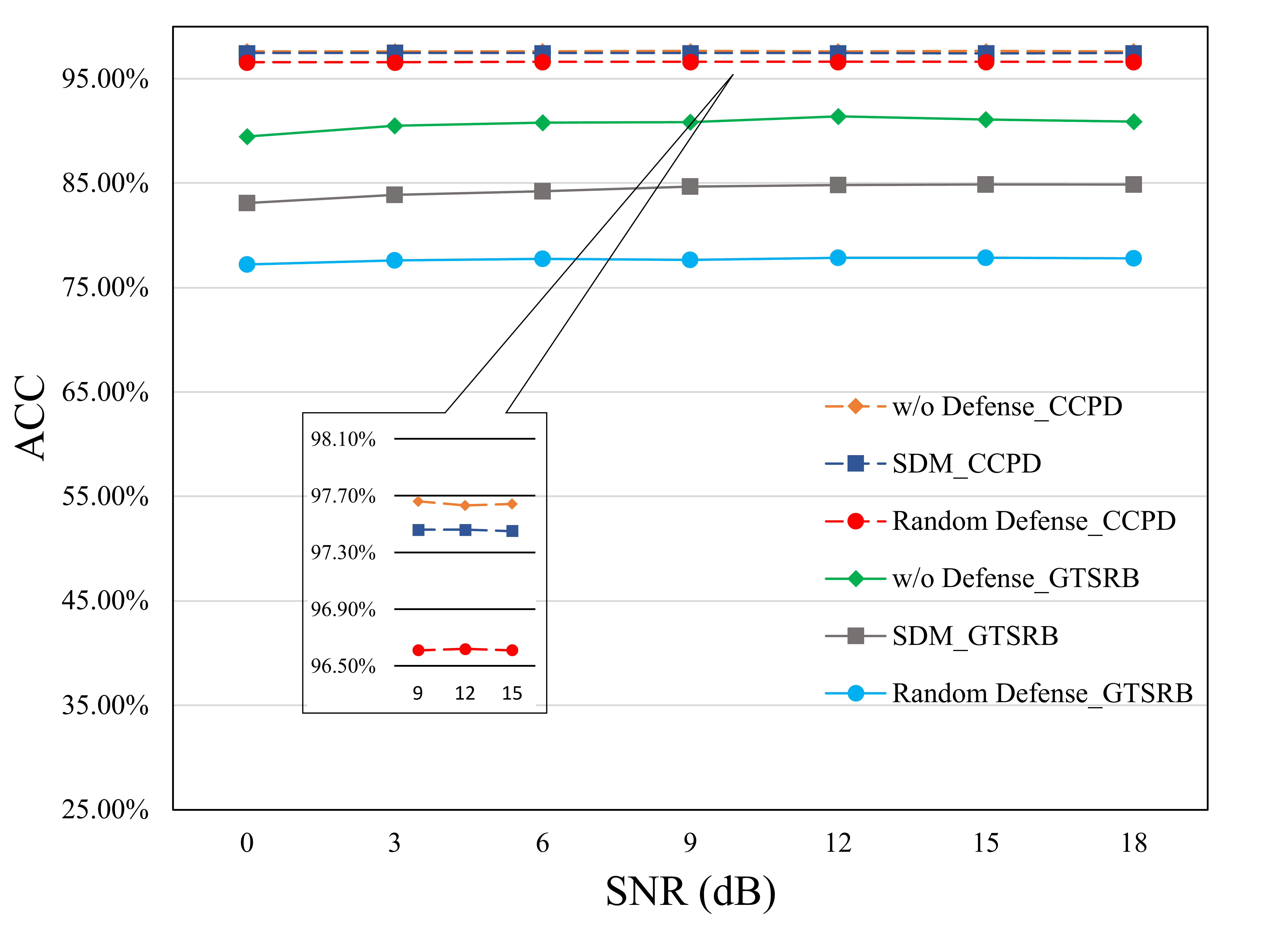}
\caption{Accuracy versus the SNR on the benign test set}
\label{fig4}
\end{figure}

		Figure \ref{fig4} shows the natural accuracy of different models when tested with benign data. Specifically, in the task of recognizing traffic signs, the SDM system exhibits a natural accuracy loss of approximately $5\%$, whereas the random defense approach results in a loss of $13\%$. Therefore, compared to random defense, SDM provides greater security while minimizing the reduction in natural accuracy.

	\section{Future Direction}
	    
		\subsection{Privacy Protection Schemes for Vehicular Metaverses}
		    
            In vehicular metaverses, the communication link between vehicles and RSUs forms a dynamic topological network structure due to vehicle mobility. During wireless communication, there is a risk of data eavesdropping. A malicious party could use eavesdropped data to learn private information such as user location and driving habits. Although there exists research on eavesdropping, the applicability of the defense schemes in dynamic environments has not been tested. Therefore, it is necessary to design a privacy protection scheme for the dynamic vehicular metaverse with changing network topology.
        
        \subsection{Customized Attack Methods for Vehicular Metaverses}
        
		    Security is a major concern for both communication systems and vehicle applications. In vehicular metaverses, however, this issue has not received enough attention. A lot of work associated with adversarial attacks, poisoning attacks, and backdoor attacks focuses on deep learning. Moreover, a targeted attack in that an attacker can control the information received by the vehicle rather than just obfuscation is more harmful. If these attacks are carried out, they are bound to raise security concerns in vehicular metaverses. Therefore, diverse and stronger attack methods for SemCom systems need to be invented to support robustness research.
		
		\subsection{Efficient and Controllable Defenses for Attacks}
		
		    This article presents a basic defense solution, which trains a semantic encoder with the capability of extracting correct semantic information from adversarial samples. However, the precision loss caused by the defensive approach should be avoided in some cases. For relatively simple pragmatic tasks, we can obtain security performance at a small cost, but for complex tasks, the cost will increase. Further research is needed to develop solutions that offer greater resistance with less cost and more controllability. The research on attacks and defenses can complement each other to further improve the security of semantic communication in vehicular metaverses.
		
	\section{Conclusions}
		Considering the data characteristic and service requirements, we have developed a hierarchical framework for vehicular metaverses that addresses the challenges of large data volumes and strict latency requirements in vehicular metaverses. To minimize data transfer, we incorporated SemCom into the framework, while also raising security concerns with an adversarial attack method specific to vehicular metaverses. To protect against such attacks, we proposed a training method for the SemCom system that can withstand semantic noise. Our evaluation showed that our defense is effective in resisting adversarial attacks, which is crucial for ensuring a secure and immersive vehicular metaverse. We believe that applying security technologies to semantic communication in both deep learning and traditional security will provide valuable insights for creating a safe and comfortable vehicular metaverse.

    \section*{Acknowledgment}
    The work is supported by NSFC under grant No. 62102099,  U22A2054, No. 62101594,  No. 62001125,  and  the Pearl River Talent Recruitment Program under Grant  2021QN02S643, and is supported by the National Research Foundation (NRF) and Infocomm Media Development Authority under the Future Communications Research Development Programme (FCP). The research is also supported by the SUTD SRG-ISTD-2021-165, the SUTD-ZJU IDEA Grant (SUTD-ZJU (VP) 202102), and the Ministry of Education, Singapore, under its SUTD Kickstarter Initiative (SKI 20210204).

\bibliographystyle{IEEEtran}
\bibliography{ref}
\vfill

    \noindent\textbf{Jiawen Kang} (kavinkang@gdut.edu.cn) received the Ph.D. degree from the Guangdong University of Technology, China in 2018. He was a postdoc at Nanyang Technological University, Singapore from 2018 to 2021. He is currently a professor at Guangdong University of Technology, China. His research interests mainly focus on blockchain, security, and privacy protection in wireless communications and networking.
\\
\\
    \noindent\textbf{Jiayi He} (hjy99911@gmail.com) received the B.Eng. degree from Shandong University, Jinan, China, in 2021. He is currently pursuing an M.S. degree with the School of Automation, Guangdong University of Technology, Guangzhou, China. His research interests include semantic communications, security and privacy, and artificial intelligence-generated content.
\\
\\
    \noindent\textbf{Hongyang Du} (HONGYANG001@e.ntu.edu.sg) received the B.Sc. degree from Beijing Jiaotong University, Beijing, China, in 2021. He is currently working toward the Ph.D. degree with the School of Computer Science and Engineering, Energy Research Institute @ NTU, Nanyang Technological University, Singapore, under the Interdisciplinary Graduate Program. His research interests include semantic communications, reconfigurable intelligent surfaces, and communication theory. He was the recipient of the IEEE Daniel E. Noble Fellowship Award in 2022. He was recognized as an Exemplary Reviewer of the IEEE Transactions on Communications in 2021.
\\
\\
    \noindent\textbf{Zehui Xiong} (zehui\_xiong@sutd.edu.sg) is currently an Assistant Professor with the Pillar of Information Systems Technology and Design, Singapore University of Technology Design, Singapore. His research interests include wireless communications, network games and economics, blockchain, and edge intelligence.
\\
\\
    \noindent\textbf{Zhaohui Yang} (yang\_zhaohui@zju.edu.cn) is currently a ZJU young professor with College of Information Science and Electronic Engineering Zhejiang Key Lab of Information Processing Communication and Networking, Zhejiang University, and also a research scientist with Zhejiang Lab. His research interests include joint communication, sensing, and computation, and semantic communication.
\\
\\
    \noindent\textbf{Xumin Huang} (huangxumin@gdut.edu.cn) received the Ph.D. degree in control science and engineering from the Guangdong University of Technology, Guangzhou, China, in 2019. From 2020 to 2021, he was a Visiting Scholar with the State Key Laboratory of Internet of Things for Smart City, University of Macau, Macau, China. He is currently an Associate Professor with the School of Automation, Guangdong University of Technology. His research interests include network performance analysis, simulation, and enhancement in wireless communications and networking.
\\
\\
    \noindent\textbf{Shengli Xie} (shlxie@gdut.edu.cn) received the M.S. degree in mathematics from Central China Normal University, Wuhan, China, in 1992, and the Ph.D. degree in automatic control from South China University of Technology, Guangzhou, China, in 1997.,He was a Vice Dean of the School of Electronics and Information Engineering, South China University of Technology from 2006 to 2010. He is currently the Director of the Institute of Intelligent Information Processing and the Guangdong Key Laboratory of Information Technology for the Internet of Things, and also a Professor with the School of Automation, Guangdong University of Technology, Guangzhou. He has authored or coauthored four monographs and more than 100 scientific papers published in journals and conference proceedings, and was granted more than 30 patents. His research interests broadly include statistical signal processing and wireless communications, with an emphasis on blind signal processing and Internet of Things.

\end{document}